\documentclass[12pt]{article}
\pdfoutput=1
\usepackage{amsmath}
\usepackage{amsfonts}
\usepackage{setspace}
\usepackage{subcaption}
\usepackage{enumerate}
\usepackage{graphicx}
\usepackage[hidelinks]{hyperref} 
\numberwithin{equation}{section}
\usepackage[utf8]{inputenc}
\newcommand{\gsim}{\lower.7ex\hbox{$\;\stackrel{\textstyle>}{\sim}\;$}}
\newcommand{\lsim}{\lower.7ex\hbox{$\;\stackrel{\textstyle<}{\sim}\;$}}
\def\O{{\mathcal O}}

\newcommand{\be}{\begin{equation}}
\newcommand{\ee}{\end{equation}}
\newcommand{\bea}{\begin{eqnarray}}
\newcommand{\eea}{\end{eqnarray}}

\newcommand{\comment}[1]{}

\newcommand{\Cint}{C\kern-1em\int}

\def\d{\partial}
\def\vphi{\varphi}

\def\O{\mathcal{O}}

\begin{document}
\vspace*{-1. cm}
\begin{center}
{\bf \Large A comment about the cosmology on a bubble wall}
\vskip 1cm
{{ Mehrdad Mirbabayi}}
\vskip 0.5cm
       {\normalsize {\em International Centre for Theoretical Physics, Trieste, Italy}}
\end{center}

\vspace{.8cm}
{\noindent \textbf{Abstract:}  
The interface between a big bubble of true AdS$_D$ vacuum expanding inside a false AdS$_D$ vacuum is a model of an inflating $D-1$ dimensional universe. It looks like an interesting setup to study fundamentals of inflation. A recent computation shows that the prediction of this model for the wavefunction of the universe disagrees with that of Hartle and Hawking. We show that this discrepancy is because the effective $D-1$ dimensional description of the model is spin-$0$ (Nordstr\"om) gravity rather than spin-2 (Einstein) gravity. 
\vspace{0.3cm}
\vspace{-1cm}
\vskip 1cm
\section{Introduction and Conclusion}
The Coleman-De Luccia (CDL) instanton is maximally symmetric \cite{CDL}. In $D$ Euclidean dimensions, it is $O(D)$ symmetric. We are particularly interested in the ``thin wall'' limit, in which the two vacua are separated by a $D-1$-sphere. In Lorentzian signature, this becomes a $D-1$-hyperboloid, which is the same as the dS$_{D-1}$ spacetime. 

This hyperbolic wall could be interpolating two AdS$_D$ vacua, with inverse curvature lengths $k_->k_+>0$. In that case, if the wall tension is close to a critical value $\sigma_c$, the dS curvature lengths $\ell$ can be very large, $ k_+ \ell\gg 1$. An interesting idea is to consider observers (and matter fields) who are localized on such a bubble wall, and view this setup as a UV completion for a $D-1$ dimensional de Sitter cosmology \cite{Banerjee18,Banerjee,Danielsson}.  In $D=5$, these authors identify the effective lower dimensional gravitational constant with
\be
\kappa_4 = \frac{2 k_+ k_-}{k_-- k_+} \kappa_5,
\ee
where $\kappa_5 \equiv 8\pi G_5$ is the bulk gravitational constant. The effective $4d$ Friedmann equation is (in the limit $k_--k_+ \ll k_+$)
\be\label{Fr}
\frac{\dot a^2 + 1}{a^2} = \frac{1}{\ell^2}\approx \frac{\kappa_4}{3} \left(3\frac{k_--k_+}{\kappa_5} - \sigma\right).
\ee
The critical tension is $\sigma_c =3(k_--k_+)/\kappa_5$, above which the false vacuum cannot decay. At least not via the CDL instanton.

When decay is allowed, the CDL instanton looks as a higher dimension embedding of the Hartle-Hawking (HH) no-boundary instanton \cite{HH}. The bubble-wall is a complex manifold that can be thought of as the expanding phase of dS$_{4}$ glued to a $4d$ hemisphere at the moment of reflection symmetry. Nevertheless, the regularized CDL instanton was found in \cite{Danielsson} to give for the leading exponent
\be\label{psiHH}
|\Psi|^2 \sim e^{-8\pi^2 \ell^2/\kappa_4},
\ee
which has the opposite sign with respect to the HH exponent.

This discrepancy might appear insignificant given that Hartle and Hawking's result is a saddle point approximation to a path integral, which is anyway defined up to an overall normalization factor. However, the prediction is unambiguous if we are dealing instead of a cosmological constant with an inflaton potential. In that case, HH wavefunction prefers larger $\ell$, and as a result a shorter period of inflation to follow after the birth of the universe. 

Let us therefore suppose that the two AdS$_5$ vacua belong to a richer multifield landscape in which the wall tension can continuously and slowly vary. Then, there can be various bubble sizes and the wavefunction \eqref{psiHH} prefers inflation to begin with smaller curvature length. So the disagreement with HH is real. The subsequent classical evolution is even more puzzling since the inflaton potential, which we naturally expect to replace $\sigma- \sigma_c $, appears with the wrong sign in the Friedmann equation \eqref{Fr}: As the inflaton rolls down, the expansion rate increases.

On the other hand, both of these signs are the correct ones in a $4d$ theory with scalar-mediated gravity rather than Einstein gravity. Let us define a metric $g_{\mu\nu}$ via
\be\label{chi}
ds^2 = e^{2\chi} (-d\tau^2 + d\Omega^2),
\ee
where $d\Omega^2$ is the metric of a 3-sphere. Consider the following action
\be\label{S4}
S[\chi,\psi] = -\frac{1}{2\kappa_4}\int d^4 x \sqrt{-g} R + S_m[\psi ,g_{\mu\nu}],
\ee
where $\psi$ stands for all matter fields (confined to $4d$), and their action $S_m$ is defined on the manifold given by \eqref{chi}. The minus sign before the first term ensures that high frequency fluctuations of $\chi$ have the correct sign kinetic term:
\be
S =-\frac{3}{\kappa_4} \int d^4x \sqrt{-g} g^{\mu\nu} \d_\mu\chi \d_\nu\chi + \cdots
\ee
The nonlinear equation of motion for $\chi$ is the same as the trace of Einstein equation, but with the opposite sign
\be\label{chieq}
R = \kappa_4 T^\mu_\mu,
\ee
which on a homogeneous and isotropic background can be integrated to give
\be
\frac{1+ \dot a^2}{a^2 } = -\frac{\kappa_4}{3} \rho + \frac{c}{a^4},
\ee
which is compatible with \eqref{Fr}. The integration constant $c$ reminds us that scalar gravity decouples from radiation (or any conformal source).

We can also apply the Hartle-Hawking proposal to this theory. For instance, with a {\em negative} cosmological constant $\Lambda<0$, the no-boundary geometry is identical to the original one with {\em positive} cosmological constant $-\Lambda$. The radius of curvature is
\be
\ell^2 = \frac{3}{\kappa_4 |\Lambda|}.
\ee
On this complex manifold, the scale factor in \eqref{chi} behaves as
\be
e^{2\chi} = - \coth \tau,
\ee
where $\tau$ starts from $i\pi/2$ at the south pole of the hemisphere, runs to $i\frac{\pi}{2}-\infty$ at the equator, then it connects to the expanding phase of dS for which $-\infty <\tau <0$. Even though this saddle is geometrically the same as the one in Einstein gravity, because of the opposite sign of the action, the leading exponent will be as in \eqref{psiHH}. 

Below we will confirm this interpretation by a bulk computation of the interaction between two low-energy sources localized on the bubble. The result will match a force mediated by a massless $4d$ scalar rather than a $4d$ graviton. This result is not new. In the context of holography and Randall-Sundrum (RS) construction, it is well known that gravity decouples from the $4d$ description when AdS$_5$ is infinitely extended in the UV \cite{Aharony}, equivalently, when the UV brane is sent to infinity in the RS model \cite{Rattazzi}. Of course, a metastable AdS vacuum is more consistently thought of as a finite bubble itself \cite{Maldacena} rather than infinite. Perhaps the only way to realize them is via inflation. Regardless of the origin, if such a bubble exists and somewhere inside a bubble of true vacuum nucleates, the discussion of the effective theory on the bubble-wall can be justified if the decay rate is small. The transition in the bulk can be thought of as the spontaneous breaking of conformal symmetry. The fine-tuning of the brane-tension is equivalent to the fine-tuning of the cosmological constant needed to obtain an ``unnaturally light dilaton'' \cite{light_dilaton}. 
\section{Derivation}
We will set up the problem in $D$ bulk dimensions. An effective $D-1$ description arises for low-energy matter fields localized on the bubble wall if its size and curvature length is much larger than the AdS lengths on both sides. Therefore, we can work in the approximation that the wall geometry is flat. In other words, we consider matter sources with characteristic momentum $p$, such that $1/\ell \ll p\ll k_+ < k_-$. 

In this approximation the bulk metric can be gauge-fixed as
\be\label{gauge}
ds^2 = N^2 dy^2 + a^2(y) g_{\mu\nu}dx^\mu dx^\nu,
\ee
where $x^\mu$ are $D-1$ coordinates. We are also working in the thin-wall approximation and fix the bubble wall to be at $y=0$. In terms of these variables, the bulk action reads
\be
S_{\rm bulk} = \frac{1}{2\kappa_D}\int dy d^{D-1}x a^{D-1} \sqrt{-g}[N a^{-2} R(g)-2\kappa_D N V_D
+N^{-1}a^{-4}(E^2 - E_{\mu\nu}^2)]
\ee
where $R(g)$ is the scalar Ricci of $g_{\mu\nu}$,
\be
E_{\mu\nu} = aa' g_{\mu\nu} + \frac{1}{2} a^2 g'_{\mu\nu},
\ee
the Greek indices are raised by $g^{\mu\nu}$, the inverse of $g_{\mu\nu}$, and prime denotes $d/dy$. The bulk potential takes two values
\be
V_D(y) =-\frac{1}{2}(D-1)(D-2) \left\{\begin{array}{cc}& k_+^2 ,\qquad y>0\\[10 pt]& k_-^2 ,\qquad y<0\end{array}\right.
\ee
The wall action consists of a tension $\sigma$ that supports the background solution, plus matter perturbations that source metric perturbations $h_{\mu\nu}$, and at leading order in the source can be parameterized as
\be
S_{\rm wall}=\frac{1}{2}\int dy d^{D-1}x a^{D-1}\sqrt{-g}(-2\sigma + h_{\mu\nu} T^{\mu\nu}) \delta(y).
\ee
The equations of motion arising from varying with respect to $N$ and $g_{\mu\nu}$ are respectively
\be\label{Neq}
a^{-2} R -2\kappa_D V - N^{-2}a^{-4}(E^2 - E_{\mu\nu}^2) =0,
\ee
and 
\be\label{geq}
\begin{split}
&\frac{1}{a^{D-3}\sqrt{- g}}[a^{D-5}\sqrt{- g} N^{-1}(E^{\mu\nu}- g^{\mu\nu} E)]'
-\frac{1}{2} a^{-4} N^{-1}  g^{\mu\nu}(E_{\alpha\beta}^2 -E^2)
+2 a^{-4} N^{-1} (E^{\mu \alpha} E_\alpha^\nu - E^{\mu\nu} E)\\[10pt]
&- N a^{-2} G^{\mu\nu} + a^{-2}(\nabla^\mu \nabla^\nu N - g^{\mu\nu} \nabla^2 N)
- \kappa_D N g^{\mu\nu} V - \kappa_D \delta(y)[\sigma g^{\mu\nu}-T^{\mu\nu}]=0,
\end{split}
\ee
where $G^{\mu\nu} = R^{\mu\nu}-\frac{1}{2}g^{\mu\nu} R$.
\subsection{Background}
First, we rederive the background solution. 
At zeroth order in perturbations, \eqref{Neq} reads
\be
(D-1)(D-2)(a'/a)^2 = -2\kappa_D V + a^{-2} \bar R,
\ee
where $\bar R$ is the uniform curvature of the constant-$y$ slices. For flat slicing, we have (after gauge fixing $N=1$ and $a(0) =1$) 
\be
a(y) =\left\{\begin{array}{cc}& e^{k_+ y} ,\qquad y>0\\[10 pt]& e^{k_- y} ,\qquad y<0.\end{array}\right.
\ee
The wall tension enters \eqref{geq}, which at the background level reduces to
\be
(2-D) a^{3-D}(a^{D-4} N^{-1} a')'+\text{lower derivative} =  \kappa_D \sigma \delta(y).
\ee
Tension has to be tuned in order to have flat slicing:
\be
\sigma = (D-2)\frac{k_- - k_+}{\kappa_D}.
\ee
Of course, the bubble wall has a small intrinsic curvature, and $\sigma$ is slightly detuned. 
\subsection{Perturbations}
A strategy to find the effective $D-1$ dimensional description is to consider $T_{\mu\nu}$ as an external source for the metric perturbations with momentum $p^\mu$ and find the generating functional $W[T_{\mu\nu}]$ at quadratic order in the source. The poles at $p^2 =0$ identify the massless degrees of freedom. At the Gaussian level, we have
\be\label{W2}
W^{(2)}[T_{\mu\nu}] = \frac{1}{4} \int d^{D-1}x h_{\mu\nu}(y=0) T_{\mu\nu},
\ee
where $h_{\mu\nu}= g_{\mu\nu}-\eta_{\mu\nu}$ is the solution to the linearized bulk equations in terms of the source $T_{\mu\nu}$.

Expanding to first order, \eqref{Neq} gives ($\pm$ correspond to the two sides)
\be\label{phi}
(D-2) k_\pm h' - 2(D-1) (D-2)k_\pm ^2 \vphi  - a^{-2} R^{(1)} =0,
\ee
where we defined 
\be
N = 1+\vphi,\qquad h= \eta^{\mu\nu} h_{\mu\nu}.
\ee
The linearized Ricci scalar is 
\be
R^{(1)} = \d^\mu\d^\nu h_{\mu\nu} - \d^2 h,
\ee
where the indices are raised by $\eta^{\mu\nu}$. Expanding \eqref{geq} to linear order and using the background solution gives
\be\label{h}
\begin{split}
&\frac{a^{1-D}}{2}[a^{D-1}(h_{\mu\nu}'-\eta_{\mu\nu} h')]' - a^{-2} G^{(1)}_{\mu\nu}\\[10pt]
&+(D-2) H a^{1-d}(a^{d-1} \vphi)'\eta_{\mu\nu} 
+a^{-2}(\d_\mu\d_\nu -\eta_{\mu\nu} \d^2)\vphi = -\kappa_D T_{\mu\nu} \delta(y),
\end{split}
\ee
and the Einstein tensor is given to first order in $h_{\mu\nu}$ by $G_{\mu\nu}^{(1)} = R^{(1)}_{\mu\nu} - \frac{1}{2} \eta_{\mu\nu} R^{(1)}$ where
\be
R^{(1)}_{\mu\nu} = -\frac{1}{2}(\d^2 h_{\mu\nu}-\d_\mu \d^\rho h_{\nu \rho} - \d_\nu \d_\rho h_{\mu\rho}+
\d_\mu\d_\nu h).
\ee
The goal is to solve \eqref{phi} and \eqref{h} for a low momentum source, with $p^2 \ll k_+^2$. We extract the transverse-traceless (tensor) component of $h_{\mu\nu}$ by decomposing 
\be
h_{\mu\nu}= h^L_{\mu\nu} + \gamma_{\mu\nu}, \qquad \gamma_{\mu}^\mu =0= \d^\mu \gamma_{\mu\nu}.
\ee
$h^L$ and $\gamma_{\mu\nu}$ couple, respectively, to the trace $T\equiv T^\mu_\mu$ and transverse-traceless part of the source $\hat T_{\mu\nu}$. In momentum space, we can decompose
\be\label{Tdec}
T_{\mu\nu} = \frac{1}{D-2} \Pi_{\mu\nu} T + \hat T_{\mu\nu},
\ee
where 
\be
\Pi_{\mu\nu} = \eta_{\mu\nu} - \frac{p_\mu p_\nu}{p^2}.
\ee
Inverting \eqref{Tdec} gives
\be
\hat T_{\mu\nu} = \frac{1}{2}\left[\Pi_{\mu\rho} \Pi_{\nu\sigma}+\Pi_{\nu\rho} \Pi_{\mu\sigma}
-\frac{2}{D-2}\Pi_{\mu\nu} \Pi_{\rho\sigma}\right]T_{\rho\sigma}.
\ee
\subsubsection{Tensor mode}
In momentum space, the equation for $\gamma_{\mu\nu}$ is
\be\label{gameq}
a^{1-D}(a^{D-1} \gamma_{\mu\nu}')' -\frac{p^2}{a^2} \gamma_{\mu\nu} =- 2 \kappa_D \hat T_{\mu\nu}\delta(y),
\ee
which imposes a jump
\be\label{gamjump}
\gamma'_{\mu\nu}(0^-)-\gamma'_{\mu\nu}(0^+) = 2\kappa_D \hat T_{\mu\nu}.
\ee
The general bulk solution on each side of the wall (indicated by $+$ for $y>0$ and $-$ for $y<0$) can be written as
\be
\gamma_{\mu\nu} = c^\pm_{\mu\nu} f^\pm_g(y) + d^\pm_{\mu\nu} f^\pm_d(y),
\ee
where $f_g$ and $f_d$ are the two independent modes. In the superhorizon limit $p\ll a(y) k_\pm$
\be
f_g^\pm(y) = 1 - \frac{p^2}{2(D-3) k_\pm^2 a^2}+\O(p^4/a(y)^4k_\pm^4),\qquad 
f_d^\pm(y) = a(y)^{1-D} (1+ \O(p^2/a(y)^2k_\pm^2)).
\ee
In the interior, the regularity in the $y\to -\infty$ limit fixes
\be
d_{\mu\nu}^- = -\frac{\Gamma\left(\frac{3-D}{2}\right)}{\Gamma\left(\frac{D+1}{2}\right)} 
\left(\frac{p^2}{4 k_-^2}\right)^{\frac{D-1}{2}} c_{\mu\nu}^-.
\ee
In the exterior, since there is no source at $y> 0$, the solution is purely decaying:
\be
c_{\mu\nu}^+ = 0.
\ee
Using the continuity of $\gamma_{\mu\nu}$ at $y=0$ and the jump \eqref{gamjump} in the derivative, we find
\be
c^-_{\mu\nu} = \frac{2\kappa_D}{(D-1)k_+} \hat T_{\mu\nu} (1+ \O(p^2/k_+^2)).
\ee
\subsubsection{Scalar mode}
Next, we take the trace of \eqref{h} to find
\be\label{ht}
\frac{1}{2}(2-D)a^{1-D}(a^{D-1} h')' +(D-1)(D-2) a^{1-D}(a^{D-2} a' \vphi)'
+\frac{D-3}{2}a^{-2} R^{(1)} +(D-2)a^{-2} p^2 \vphi = -\kappa_D T\delta(y).
\ee
Noting that $a'/a$ is discontinuous at $y=0$, we obtain the following jump condition,
\be\label{hjump}
(D-2)[h'(0^-)-h'(0^+)] - 2(D-1)(D-2) (k_--k_+) \vphi(0) =-2\kappa_D T.
\ee
To calculate $W^{(2)}[T_{\mu\nu}]$, we only need $h(0)$. This can be determined without solving for the full $h(y)$. First, we use equations \eqref{phi} and \eqref{hjump} to find
\be\label{R}
R^{(1)}(y=0) = 2\kappa_D \frac{k_+ k_-}{k_- - k_+} T.
\ee
Next we use the residual gauge symmetry in \eqref{gauge} that allows setting $\d^\mu h_{\mu\nu} =0$ at one point, which we choose $y=0$. With this choice
\be
R^{(1)}(y=0) = p^2 h(y=0).
\ee
So we can solve for $h(y=0)$ using \eqref{R}. 
\subsubsection{Massless spectrum}
Substituting the solution for $h_{\mu\nu}(0)$ in \eqref{W2} gives
\be\label{W}
W^{(2)}[T_{\mu\nu}]=- \frac{\kappa_D k_- k_+}{2(D-2)(k_- - k_+)}\int d^{D-1} x\  T\frac{1}{\Box} T + {\rm local}.
\ee
As anticipated, the KK zero mode of graviton has decoupled, but there is a massless scalar. Matching this to $D-1$ dimensional version of \eqref{S4} gives the same gravitational constant as expected from the (sign reversed) Friedmann equation
\be
\kappa_{D-1} = \frac{(D-3) k_+ k_-}{k_-- k_+} \kappa_D.
\ee
\section*{Acknowledgments}
I thank Lorenzo Di Pietro, Victor Gorbenko, Oliver Janssen and Giovanni Villadoro for useful discussions. 
\bibliography{bibwall}

\end{document}